\documentclass[preprint, 12pt]{elsarticle}

\usepackage{amssymb}
\usepackage{lipsum}
\usepackage{lineno}

\usepackage{siunitx}
\usepackage{graphicx}
\usepackage{color}
\usepackage{xcolor}
\usepackage[colorlinks = true, linkcolor = blue, urlcolor  = blue, citecolor = blue, anchorcolor = blue]{hyperref}
\usepackage{mathtools}
\usepackage{orcidlink}
\usepackage{bm}

\journal{Nuclear Physics A}

\begin{document}

\begin{frontmatter}


\title{Aperture limitation localization using beam position and beam loss monitor measurements}


\author[binp,nstu]{Yu.~Maltseva\corref{mail}\,\orcidlink{0000-0003-2753-2552}}
\ead{Yu.I.Maltseva@inp.nsk.su}
\cortext[mail]{Corresponding author.}
\author[binp,nstu,skif]{I.~Morozov\,\orcidlink{0000-0002-1821-7051}}

\affiliation[binp]{organization={Budker Institute of Nuclear Physics SB RAS},
            city={Novosibirsk},
            postcode={630090},
            country={Russia}}

\affiliation[nstu]{organization={Novosibirsk State Technical University},
            city={Novosibirsk},
            postcode={630073},
            country={Russia}}

\affiliation[skif]{organization={Synchrotron Radiation Facility "SKIF"},
            city={Koltsovo},
            postcode={630559},
            country={Russia}}


\begin{abstract}

The efficiency of beam injection in circular accelerators can be impacted by unknown aperture limitations in the vacuum chamber. These limitations can be detected by introducing localized distortions to the closed orbit of the circulating beam observed at beam position monitors (BPMs). The localized nature of introduced closed orbit distortions enables targeted identification of loss regions. In this paper, we present a method for localizing beam losses designed primarily for circular electron accelerators using a set of sequentially placed scintillator-based beam loss monitors (BLMs). By moving these monitors along the accelerator lattice, the proposed technique achieves sub-meter spatial resolution in beam loss localization. Relative calibration of the BLMs was performed using a strontium-90 radioactive source. We report experimental results from the implementation of this method at the VEPP-4M collider.

\end{abstract}

\begin{keyword}

Aperture limitation \sep Beam loss \sep Beam loss monitors \sep Photomultipliers \sep Scintillators \sep Beam position monitors

\end{keyword}

\end{frontmatter}

\tableofcontents


\section{Introduction}
\label{section:introduction}


A beam of charged particles circulates within a vacuum chamber, which is a complex system comprising numerous interconnected components, including insertion devices, probes, shutters, radiation detectors, and RF system elements. Various defects in these components can restrict the beam aperture, leading to degraded beam injection efficiency. In addition to various physical constraints (misplaced components, foreign objects), improper geodetic positioning of diagnostic systems can also manifest themselves as aperture limitations. For optimal accelerator performance, such aperture limitations must be promptly identified and mitigated.

Standard methods for detecting aperture limitations involve measuring ionizing radiation produced by lost beam particles or comparing beam current readings at different locations along the accelerator lattice. This problem is particularly critical in hadron accelerators, where beam losses can have severe consequences, such as quenching superconducting magnets or causing significant material damage. Therefore, such accelerators are equipped with extensive diagnostic systems to monitor and mitigate these risks.

To identify aperture limitations, various advanced techniques are employed at accelerator facilities. At the LHC, a combination of induced emittance blow-up and collimator scans is used, where movable collimator jaws are adjusted alongside other aperture measurement methods~\cite{martinez}. The SPS utilizes an automatic bump scan technique, which has successfully localized aperture restrictions, including non-conforming RF fingers~\cite{kain}. Additionally, another method was developed at Fermilab for the PIP-II linear accelerator~\cite{shemyakin}, employing beam modulation via AC dipole corrector pairs combined with movable scrapers. This approach enables direct observation of beam loss locations using BLMs, supplemented by indirect monitoring through beam current monitors and BPMs.

This work presents a method for identifying aperture limitations in electron circular accelerators, combining turn-by-turn (TbT) beam current measurements from stationary BPMs with beam loss detection using portable scintillator-based BLMs. These BLMs demonstrate excellent sensitivity to particle losses for beam energies above \SI{100}{\mega\electronvolt}~\cite{malts}. The technique involves locally distorting the beam orbit near suspected aperture restrictions, followed by excitation of coherent betatron oscillations via a pulsed electromagnetic kick. BPMs provide TbT beam current measurements after the kick. The above procedure can be employed only when beam circulation is present. In this case, if TbT current readings and beam excitation are available, an aperture limitation can be localized within the region enclosed by a pair of consecutive BPMs where a significant current drop is observed. 

To enhance spatial resolution in localizing aperture limitations, a set of portable scintillator-based BLMs can be deployed within the accelerator region identified by initial BPM current measurements. These detectors provide discrete sampling of the longitudinal beam loss distribution induced by the aperture limitation. Rearrangement of the BLMs enables iterative refinement of the loss location. To ensure correct comparison of signals across all monitors, prior relative sensitivity calibration of the BLM system is essential. This region can be further refined by additional numerical Monte Carlo simulations to extrapolate the obstacle location. These simulations help reduce the number of BLM adjustments and allow examination of possible aperture restrictions within accelerator components where BLMs cannot be directly placed.

\begin{figure*}[!th]
    \begin{center}
    \includegraphics[width=\linewidth,keepaspectratio]{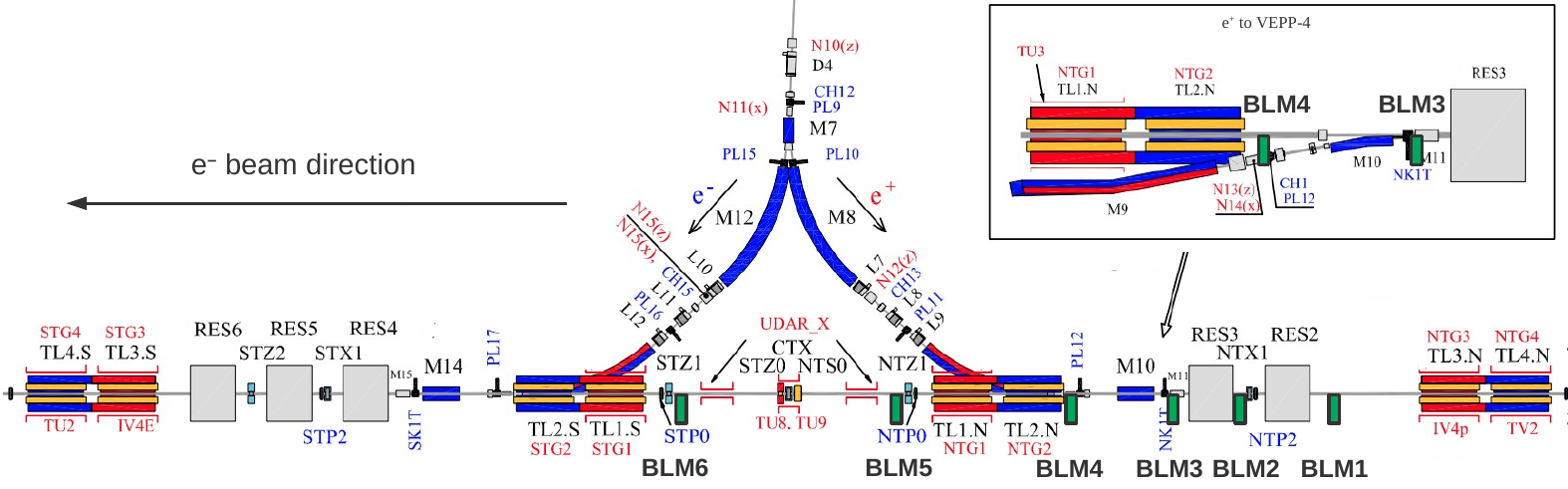}
    \end{center}
    \caption{Layout of the VEPP-4M technical region  (around $\SI{40}{\meter}$ long straight section). BLM positions are indicated by green rectangles (labeled BLM1--6). BPMs are labeled as NTP2, NTP0, STP0, and STP2. Coherent beam excitation is induced by the UDAR\_X and IV4p kickers.}
    \label{fig:layout}
\end{figure*}

To validate the proposed beam loss localization method, experimental studies were conducted at the VEPP-4 accelerator complex~\cite{vepp4m}. The VEPP-4 facility supports high-energy physics experiments, nuclear physics research, and synchrotron radiation studies. The main part of the complex is the electron-positron collider VEPP-4M, operating at beam energies up to \SI{5}{\giga\electronvolt}. The investigation was motivated by observed asymmetries in beam lifetime during induced closed orbit distortions of opposite directions in the technical region (around $\SI{40}{\meter}$). This region houses RF resonators and serves as the beam injection point. When an observable drop in the circulating beam lifetime occurred due to the introduced orbit distortion (indicating that the beam was close to the aperture restriction), coherent beam oscillations were induced around this new closed orbit, and TbT current measurements were recorded at the BPMs. A notable drop was observed between two consecutive BPMs located in the technical region, confirming the presence of an aperture restriction there. For precise localization of the aperture limitation, five identical scintillator-based BLMs were deployed. Their arrangement, along with BPM positions, is shown in Figure~\ref{fig:layout}.


The structure of this article is as follows: Section~\ref{section:bpm:localization} describes the beam current drop localization procedure using TbT data analysis. Section~\ref{section:blms} details the Monte Carlo simulation employed to model BLM response, the results of the BLM relative calibration performed using a strontium-90 radioactive source. Section~\ref{section:experiment} presents results of experimental aperture limitation measurements using both BPM and BLM systems, which are compared with the simulations for several beam loss scenarios.


\section{BPM based localization}
\label{section:bpm:localization}


BPMs are widely used as efficient, non-destructive diagnostic instruments for charged particle beams~\cite{forckBPM}. These devices measure induced charges generated by the passing beam, enabling reconstruction of the beam's center-of-mass position and other quantities of interest depending on the monitors construction and purpose. A common BPM design incorporates four in-vacuum electrodes (buttons or strips)~\cite{smaluk}. Transverse beam centroid positions can be inferred from electrode signals difference, while sum of the signals provide a beam current estimate.

The experimental studies were performed at the VEPP-4M collider,  which is equipped with 54 dual-plane button BPMs with TbT acquisition capability~\cite{beh}. However, the current measured from BPM sum signals is less accurate (\SI{5}{\percent} rms variation across monitors) than that obtained from DCCT measurements. BPM synchronization is an additional requirement for correct interpretation of acquired TbT signals~\cite{anomaly}.

As BPMs are distributed throughout the accelerator lattice, aperture limitations can be localized between consecutive monitor pairs. When the beam current drop caused by the aperture limitation exceeds BPM measurement uncertainties, a distinct step-like drop becomes observable in ordered TbT signals from all BPMs. To improve the signal-to-noise ratio between the single-turn current loss and BPM measurement accuracy, we implement the following two-step procedure.

The localization procedure begins with closed orbit bump scans to probe potential aperture restrictions. Localized orbit distortions are generated using dedicated horizontal and vertical corrector magnets. Additionally, at the VEPP-4M, the closed orbit can be globally shifted in the vertical direction within the entire technical region (Figure \ref{fig:layout}). Initial scans use extended bumps spanning multiple lattice sections between BPMs, with beam lifetime variations (measured through DCCT beam current fitting) serving as the primary indicator of potential aperture limitations.

Upon detecting a significant beam lifetime variation, coherent betatron oscillations are excited using dedicated electromagnetic kickers~\cite{kick}. The observed lifetime reduction indicates the beam is close to an aperture limitation. Subsequent excitation of coherent beam oscillations should produce localized beam loss in the examined region. Horizontal and vertical oscillations are induced by the UDAR\_X and IV4p kickers, respectively (Figure \ref{fig:layout}). When the oscillation amplitude exceeds a critical threshold, a step-like current drop becomes observable between consecutive BPMs bracketing the aperture restriction. These current drops will appear separated by several turns, depending on the transverse betatron frequencies.


\section{Simulation and experimental studies for BLM system}
\label{section:blms}


\subsection{Beam loss detection principle}


BLMs are essential diagnostic instruments for any type of particle accelerator, providing critical beam loss detection capabilities during both commissioning and routine operation. Information about the causes of the beam losses contributes significantly to machine performance optimization~\cite{witt,zhukov}. These highly sensitive detectors can register even small portions of lost particles, proving invaluable for beam threading during commissioning and accelerator tuning in general~\cite{heps}. Installed outside of the vacuum pipe at critical locations, BLMs detect secondary reaction products generated when high-energy beams interact with the vacuum chamber walls. For electron beams exceeding \SI{100}{\mega\electronvolt}, the dominant interaction mechanism is electromagnetic shower~\cite{forck}, which generates secondary electrons, positrons and photons.

\begin{figure}[!th]
    \begin{center}
      \includegraphics[width=0.5\textwidth,keepaspectratio]{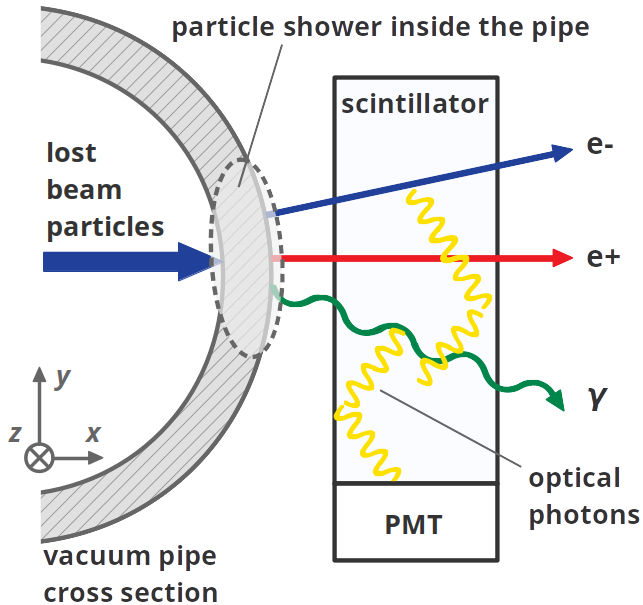}
    \end{center}
    \caption{Schematic view of the scintillator-based BLM detection principle. The beam direction is along the z axis.}
    \label{fig:principle}
\end{figure}

For aperture limitation detection, we employed scintillator-based BLMs comprising a plastic scintillator rod coupled to a photomultiplier tube (PMT). This design was selected for its fast response time, high radiation sensitivity, and manufacturing and operational simplicity. The detection principle of this BLM configuration is illustrated in Figure~\ref{fig:principle}. The detection mechanism relies on scintillation light produced when shower particles pass through the scintillator volume. This light is then guided to the PMT, where it undergoes conversion to an amplified electrical signal followed by digitization. The resulting signal is proportional to the number of detected lost beam particles.


\subsection{Monte Carlo simulations}


Monte Carlo simulations serve as a powerful statistical based technique for optimizing BLM design parameters (geometry, material composition) and positioning~\cite{blm_mc}, while also modeling expected loss patterns under various beam loss scenarios. For this study, we employed the FLUKA code~\cite{fluka} to interpret the measured BLM responses with simulated beam loss distributions and to precisely identify the aperture restriction location along the accelerator lattice.

Before performing beam loss simulations, we conducted an analysis of potential loss mechanisms. Since no known physical interventions had been made in the vacuum chamber prior to the observed losses, we could confidently exclude the presence of a sizable foreign object such as maintenance tools left inadvertently~\cite{jas14}. The observed aperture restriction likely originated from detached small vacuum components like RF fingers~\cite{fingers}, malfunctioning of insertable devices, or systematic errors in beam orbit measurements due to BPM geodetic setpoints. The latter case effectively creates artificial aperture limitations through incorrect beam position readings.

To estimate the BLM response, we performed simulations of various loss scenarios for a \SI{1.9}{\giga\electronvolt} electron beam (corresponding to injection energy) hitting the VEPP-4M vacuum chamber wall at a small incident angle relative to the equilibrium orbit. To optimize simulation runtime, we implemented several simplifications: the beam was modeled as a point source with zero angular and energy spreads, the vacuum chamber was approximated as a cylindrical pipe, and magnetic elements and RF cavities were represented as simple geometric blocks without electromagnetic fields. The process of electromagnetic shower generation when electron beam hits the metal surface of the vacuum pipe and the scintillation process inside the scintillator volume were simulated independently.

To model the secondary particles distribution, we represented the scintillator detector as a \SI{15}{\meter} long plate (spanning most of the technical region) with \SI{10}{\centi\metre} width, the latter corresponding to the actual scintillator rod width, positioned directly outside of the vacuum pipe. This geometry provides continuous distribution observation of secondary particles capable of producing scintillation light within the virtual detector along the accelerator lattice.

According to our electromagnetic shower simulations, the number of secondary photons passing through the chosen scintillator volume is six times higher than the number of charged particles. However, organic scintillators that we have used are significantly more sensitive to electrons and positrons than to photons~\cite{leo}. FLUKA simulations of scintillation light production in our specific detector geometry showed that charged particles generate 30 times more optical photons than electromagnetic radiation, making the latter contribution to the total optical signal only about \SI{20}{\percent}. Therefore, to simplified the analysis, we considered only secondary charged particles in our subsequent simulations.

\begin{figure}[!th]
    \begin{center}
      \includegraphics[width=0.7\textwidth,keepaspectratio]{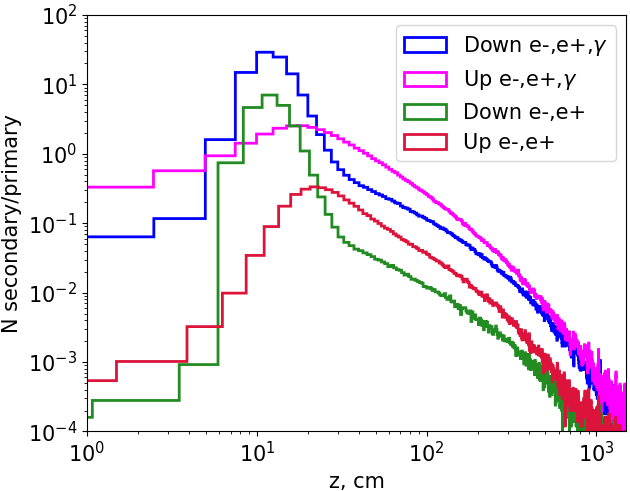}
    \end{center}
    \caption{The longitudinal distribution of secondary particles ($e-$, $e+$ and $\gamma$) participating in scintillation is shown for two detector placements: directly below the beam impact point on the vacuum pipe ("Down") and on the opposite side ("Up").}
    \label{fig:fluka1}
\end{figure}

Figure~\ref{fig:fluka1} shows the longitudinal distribution of secondary particles ($e-$, $e+$ and $\gamma$) participating in scintillation when the beam hits the vacuum pipe at the scintillator detector location or on the opposite side. The electron beam direction is along the $z$ axis and the impact point is located at $z = \SI{0}{\centi\meter}$. The obtained distributions exhibit a peak (with a FWHM of up to \SI{30}{\centi\meter}) followed by a decline region where the secondary particle distribution is nearly identical on both the impact and opposite sides. Furthermore, while secondary photons contribute to the total particle count, they do not alter the shape of the distribution.


\begin{figure}[!th]
    \begin{center}
      \includegraphics[width=0.8\textwidth,keepaspectratio]{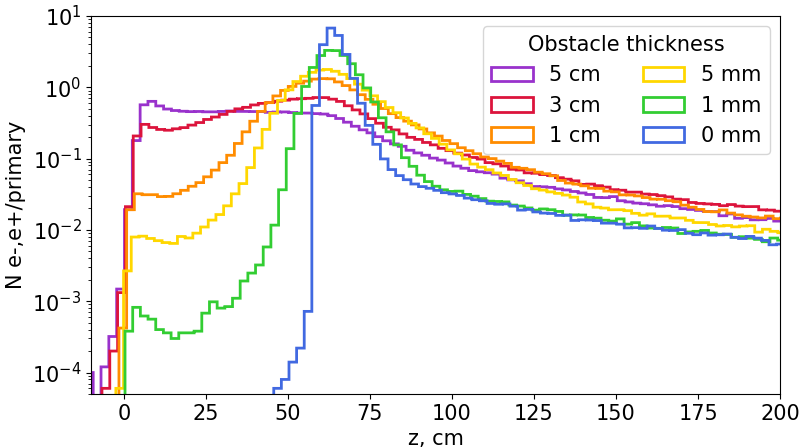}
    \end{center}
    \caption{The longitudinal distribution of secondary charged particles passing through the scintillator placed at the impact point is examined for two scenarios: beam impact on an obstacle inside the vacuum pipe with thickness of \SI{1}{\milli\meter} up to \SI{5}{\centi\meter}, and direct beam impact on the vacuum pipe wall -- as obstacle thickness of \SI{0}{\milli\meter}.}
    \label{fig:fluka2}
\end{figure}

Figure~\ref{fig:fluka2} shows the longitudinal distribution of secondary charged particles for two distinct scenarios: the beam hitting the vacuum pipe directly, and the beam impacting a metal obstacle of varying thickness inside the pipe, with thicknesses up to \SI{5}{\centi\meter}. This variation in obstacle thickness allows for a systematic investigation of secondary particle behavior and determination of the maximum achievable width of their longitudinal distribution. The scintillator detector is positioned directly below the impact point, with each histogram bin corresponding to the physical dimensions of the scintillator used in experiments, thereby representing the number of particles passing through the actual scintillator volume.

The presence of an obstacle results in a broader peak in the longitudinal distribution and an increased yield of secondary particles for high-energy beams. In all cases, the distribution displays an exponential decay at distances greater than $\SI{1.5}{\meter}$ from the impact point, with almost zero distribution observed in front of the obstacle. To ensure consistent alignment of the peaks in the analysis, the obstacle is placed at  $z = \SI{0}{\centi\meter}$ along the electron beam direction, while the beam impact on the vacuum pipe (without the obstacle) occurs at $z = \SI{53.7}{\centi\meter}$.


\begin{figure*}[!th]
    \begin{center}
      \includegraphics[width=0.7\textwidth,keepaspectratio]{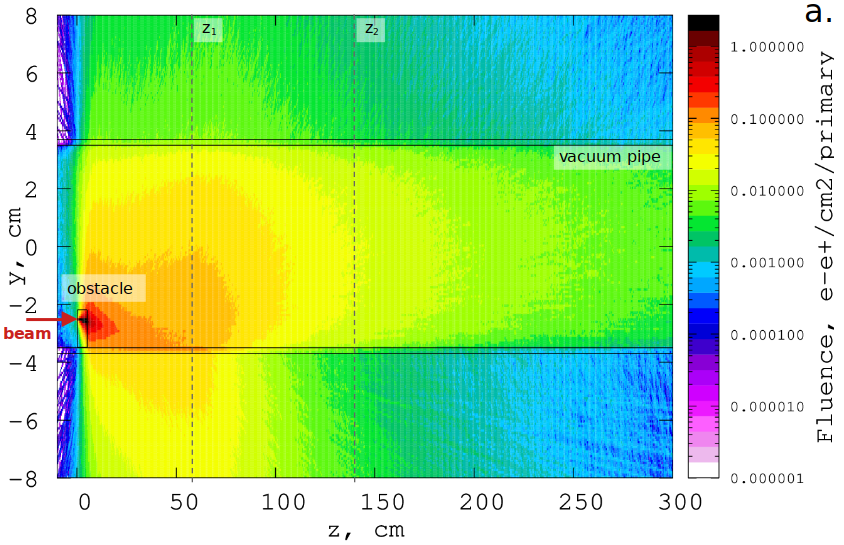}
      ~
      \vspace{0.1cm}
      \includegraphics[width=0.475\textwidth,keepaspectratio]{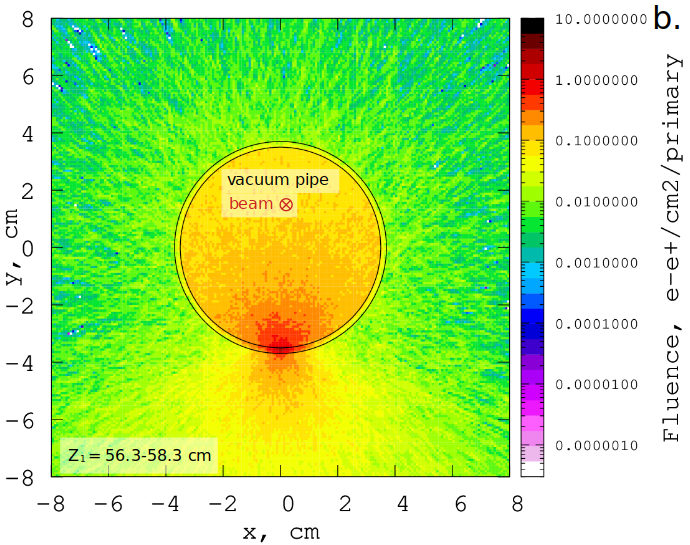}
      \includegraphics[width=0.475 \textwidth,keepaspectratio]{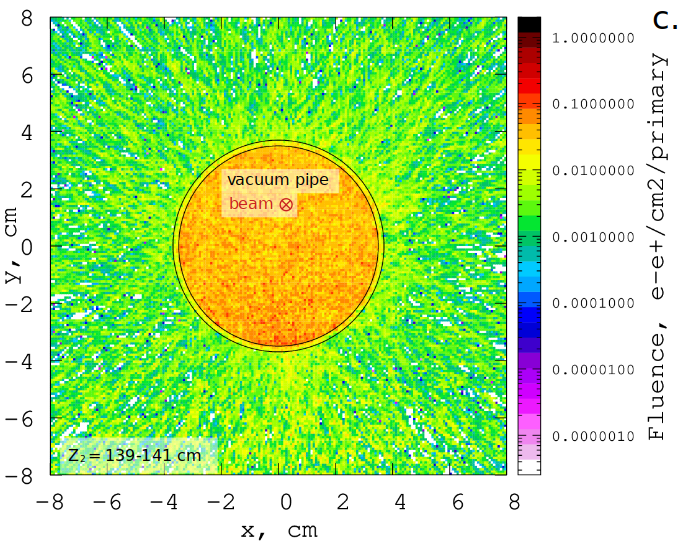}
    \end{center}
    \caption{The fluence of secondary charged particles resulting from beam loss at a \SI{5}{\centi\meter}-thick metal obstacle inside the vacuum pipe: a) in the longitudinal plane; b) in the transverse plane near the impact point at $z_1 = \SI{57.3}{\centi\meter}$; and c) the transverse plane at $z_2 = \SI{140}{\centi\meter}$ after the impact point.}
    \label{fig:fluka3}
\end{figure*}

The transverse distribution of secondary charged particles outside the vacuum pipe is also inhomogeneous. Figure~\ref{fig:fluka3} shows a characteristic distribution of secondary charged particles resulting from beam loss at an obstacle inside the vacuum pipe. The distribution shows a pronounced peak at the impact location in both the longitudinal (Figure~\ref{fig:fluka3}a) and transverse planes (Figure~\ref{fig:fluka3}b), with an almost azimuthally symmetric transverse distribution excluding the impact region ($z_1 = \SI{57.3}{\centi\meter}$). Beyond \SI{1}{\meter} ($z_2 = \SI{140}{\centi\meter}$), the transverse distribution becomes fully azimuthally symmetric across the entire pipe cross-section (Figure~\ref{fig:fluka3}c). This symmetry occurs due to backscattering of secondary particles within the vacuum pipe and their subsequent exit on the opposite side~\cite{malts}.
Consequently, the use of multiple BLMs over an extended distance allows for flexible placement of the monitors in the transverse plane, as reliable loss localization can be achieved regardless of their exact transverse positioning.


\subsection{Monitor Calibration}
\label{subsection:calibr}


In this study, a set of five identical BLMs was employed. Each BLM consists of a polystyrene-based SC-201 scintillator (IHEP, Russia) coupled to a Hamamatsu R1924A PMT. The SC-201 scintillator has a decay time of \SI{3}{\nano\second} and a light output equivalent to \SI{55}{\percent} of anthracene. Scintillator blocks are cylindrical shape with \SI{10}{\centi\meter} length and \SI{2.5}{\centi\meter} diameter, matching the PMT photocathode area. To optimize light collection, each scintillator block is wrapped in reflective Tyvek paper. The R1924A PMT was chosen due to its photocathode peak spectral response aligning with the maximum of the scintillator emission spectrum at \SI{420}{\nano\meter}, thereby enhancing the overall BLM sensitivity. To ensure consistent measurements across all BLMs, a relative calibration of their sensitivities was performed, allowing direct comparison of beam loss signals.

The performance of the PMTs and scintillators was characterized independently using a $^{90}\textrm{Sr}$ radioactive source for calibration. $^{90}\textrm{Sr}$ undergoes a two-stage $\beta^{-}$ decay cascade. First, $^{90}\textrm{Sr}$ decays into unstable $^{90}\textrm{Y}$ with the emission of an electron (up to \SI{546}{\kilo\electronvolt}), followed by the decay of $^{90}\textrm{Y}$ into stable $^{90}\textrm{Zr}$ with the emission of an electron (up to \SI{2.28}{\mega\electronvolt})~\cite{sr90}. During the calibration procedure, the radioactive source was placed at the scintillator end face, while the coupled PMT detected the scintillation light produced by the $\beta$-decay electrons.

While PMT calibration is typically performed using an LED directed at the photocathode to measure gain response, we opted for a radioisotope-based method. This approach is better aligned with the actual operating conditions, as the scintillator emits light over a broader spectrum than an LED due to $\beta^{-}$ decay electrons energy spread. And the emission spectrum closely matches the PMT's region of maximum spectral sensitivity~\cite{calibr}.

\begin{figure*}[!th]
    \begin{center}
    \includegraphics[width=0.575\textwidth,keepaspectratio]{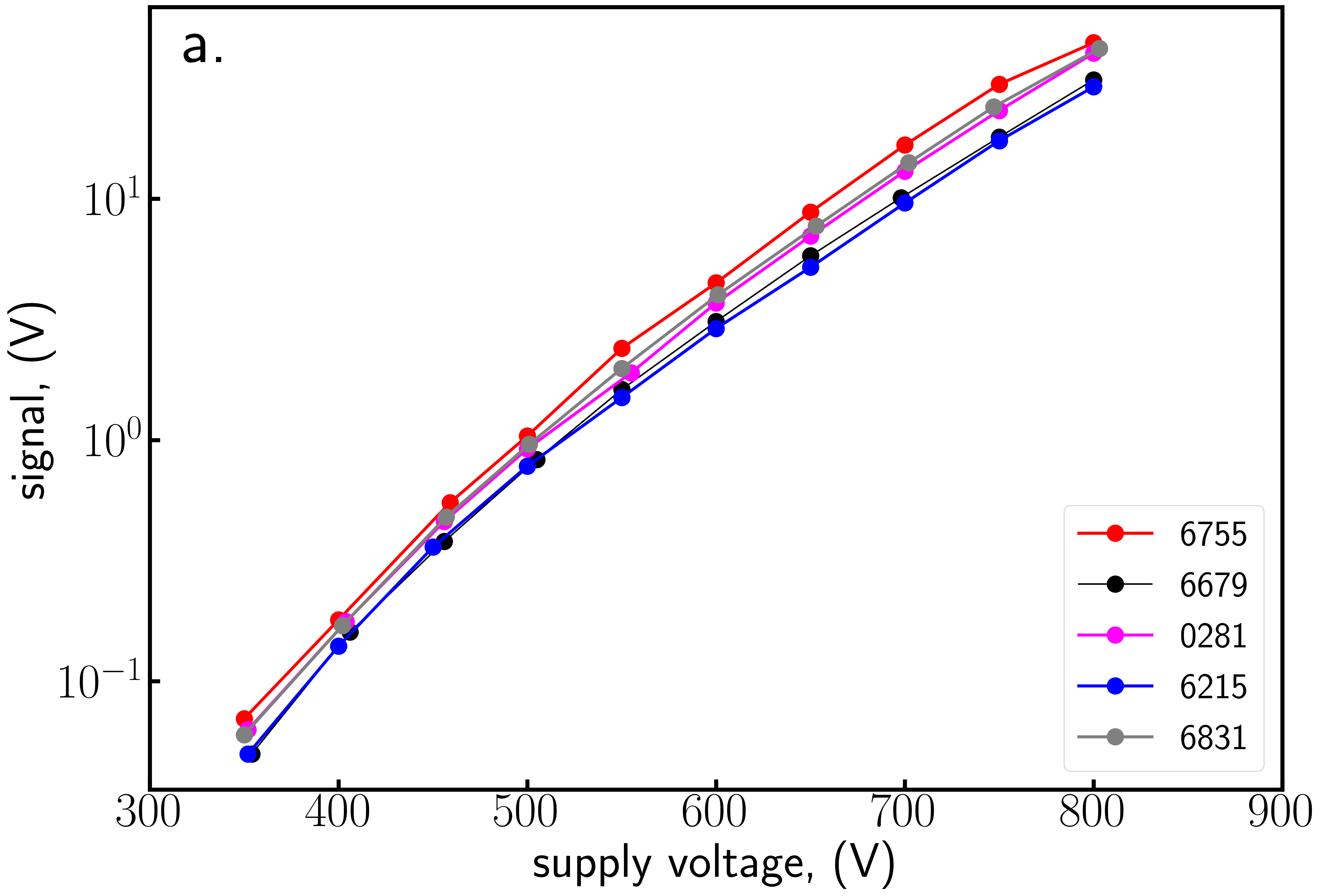}
    ~
    \vspace{0.1cm}
    \includegraphics[width=0.475\textwidth,keepaspectratio]{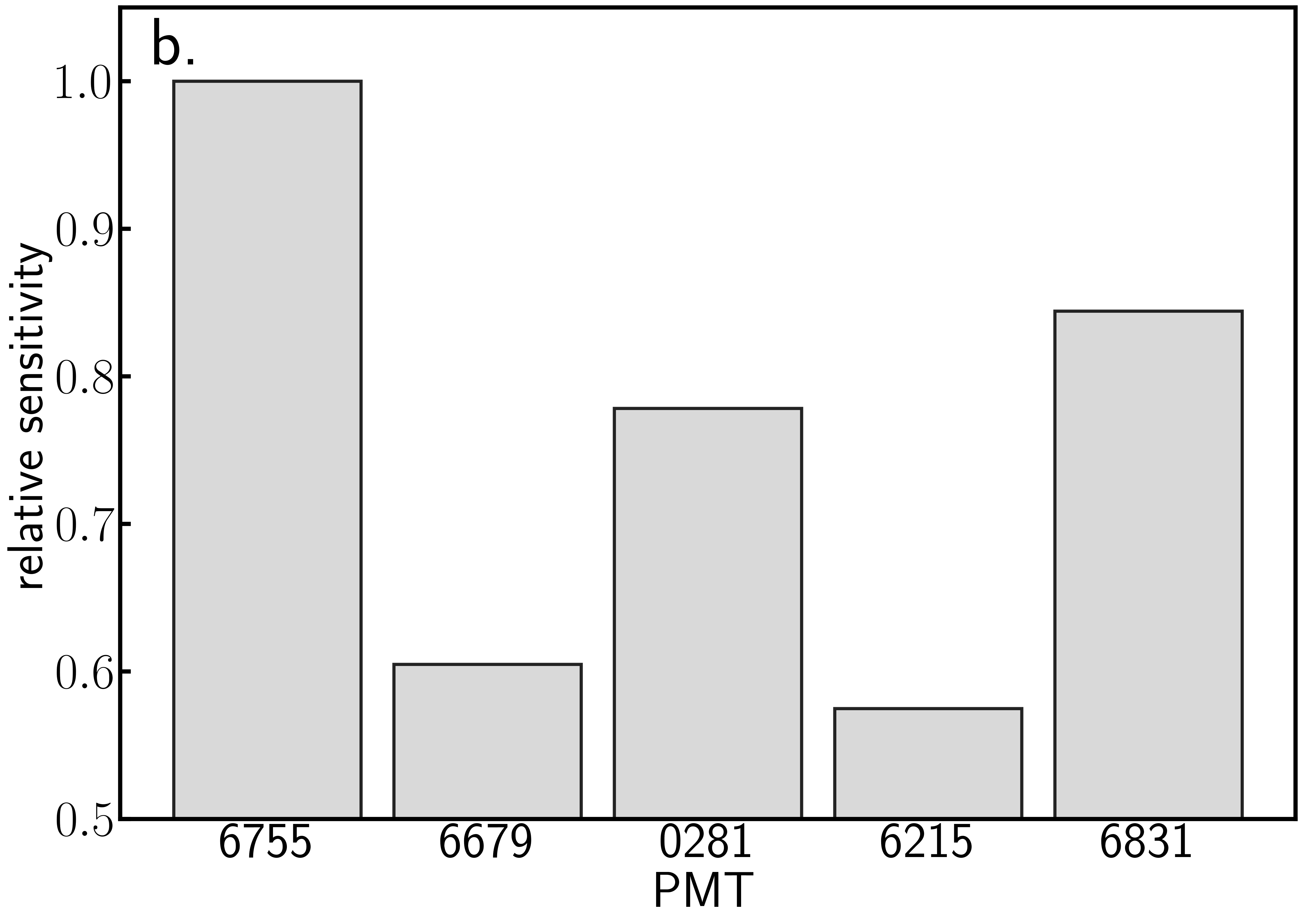}
    \includegraphics[width=0.475\textwidth,keepaspectratio]{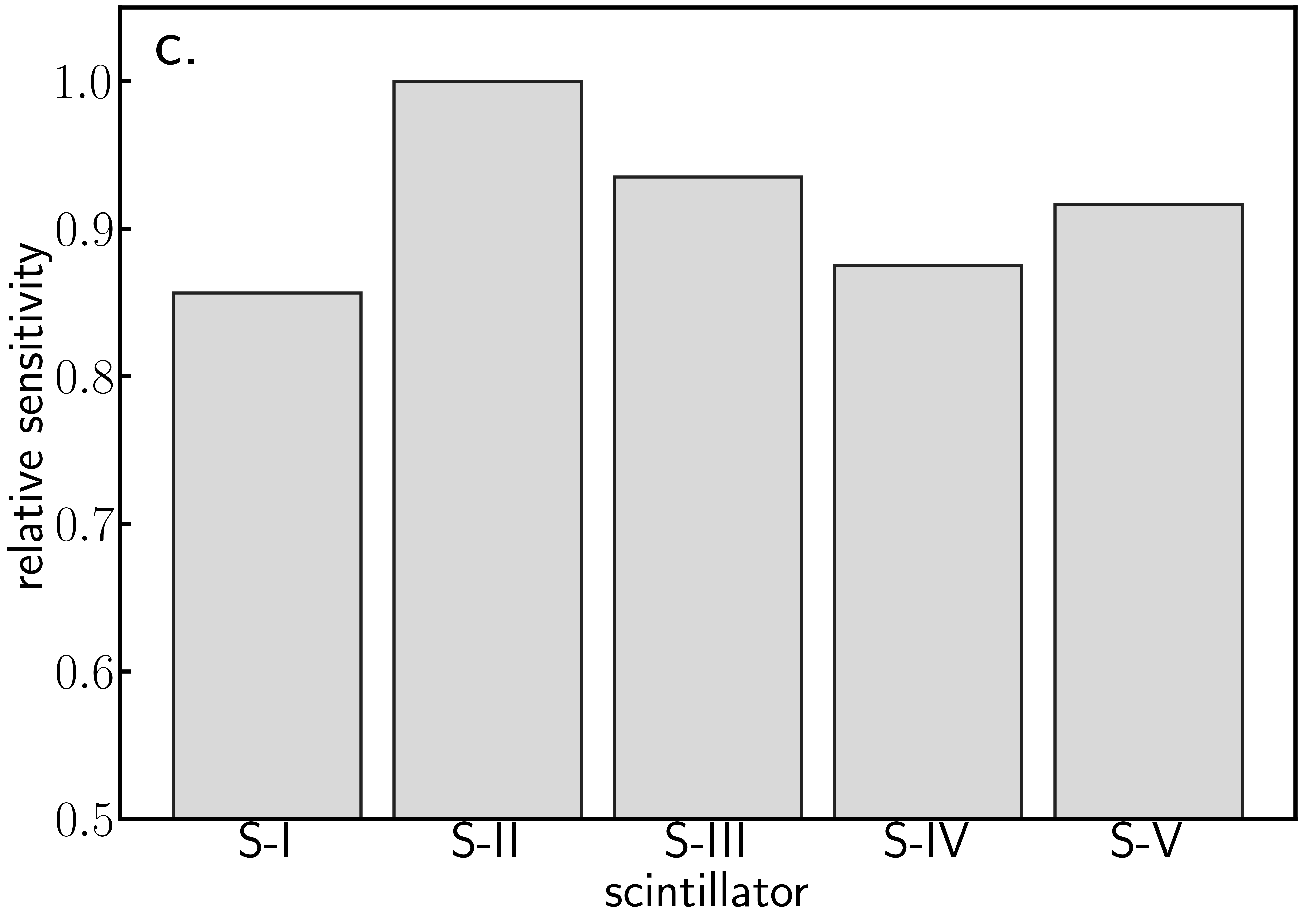}
    \end{center}
    \caption{a) PMT gain responses with respect to the supply voltage; Relative sensitivity of: b) PMTs, and c) scintillators, both measured using a radioactive $^{90}\textrm{Sr}$ source.}
    \label{fig:calibration}
\end{figure*}

For relative PMT response measurements, a single reference scintillator block was selected to record signals from all five PMTs. The gain response of each PMT was characterized as a function of supply voltage across its operational range (Figure~\ref{fig:calibration}a). The signals from PMTs have values of up to tens of volts due to the integration time of the used digital voltmeter. The operational voltage of \SI{700}{\volt} was experimentally obtained based on monitor sensitivity to observed beam losses. Relative PMT sensitivities at this voltage are shown in Figure~\ref{fig:calibration}b.

To measure the relative scintillator response, we selected a single reference PMT to record the light signals from all five scintillator blocks. As shown in Figure~\ref{fig:calibration}c, the spread in the relative light output among the scintillators was within \SI{15}{\percent} peak-to-peak. These calibration procedure ensured accurate comparison of responses across different BLMs.


\section{Experimental aperture limitation localization}
\label{section:experiment}


\subsection{TbT current measurements from BPMs}


During routine operation, a significant asymmetry in beam lifetime depending on the direction of this vertical shift (upward vs. downward) was observed. The downward orbit shifts resulted in faster beam lifetime degradation, with an observed asymmetry of approximately \SI{5}{\cm}. This behavior suggested the presence of an aperture limitation in the ring. To identify the potential aperture restriction, we conducted a systematic investigation of several candidate locations within the technical region.

\begin{figure}[!th]
    \begin{center}
      \includegraphics[width=\columnwidth,keepaspectratio]{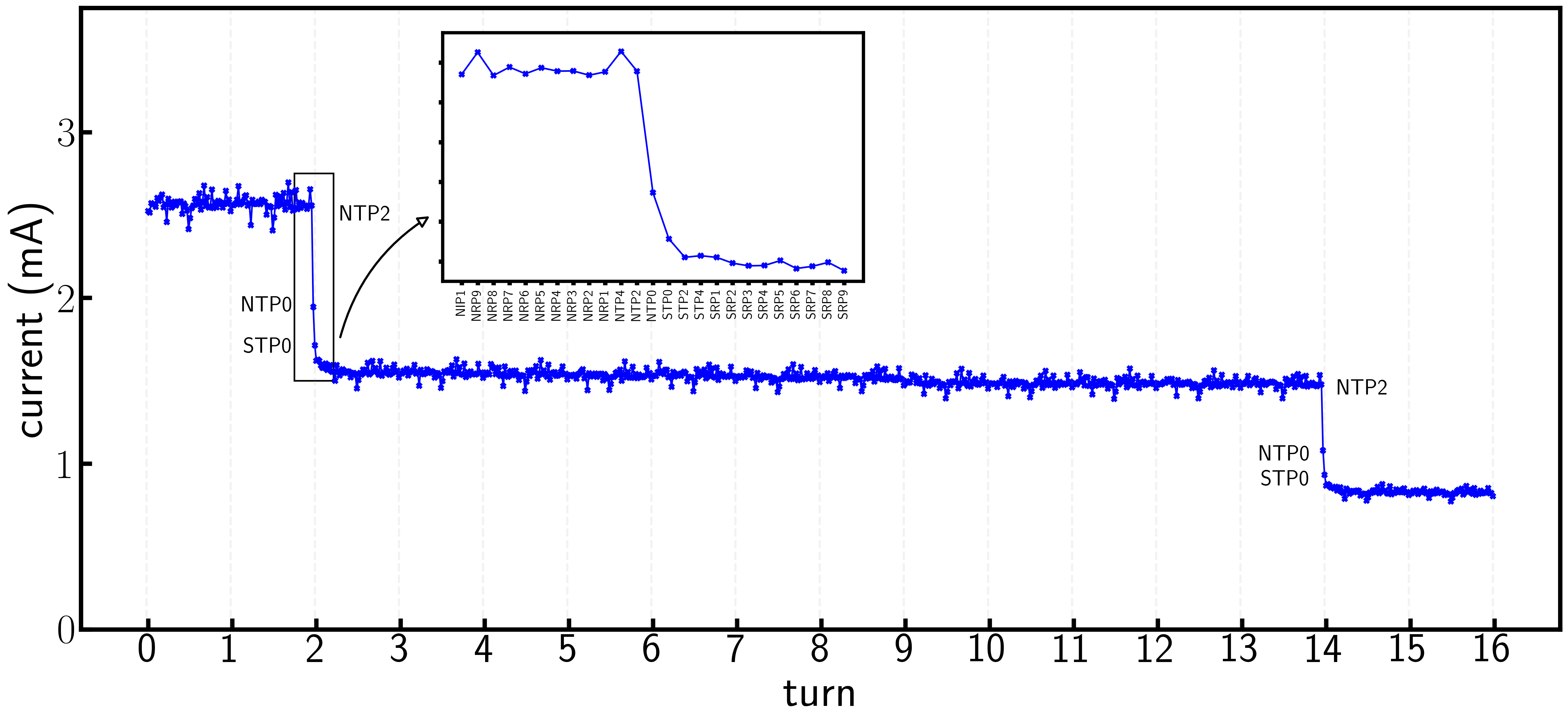}
    \end{center}
    \caption{Combined TbT beam current measurements across all BPMs during the first sixteen turns. Vertical orbit distortion is introduced in the technical region, transverse oscillations are excited by electromagnetic kick. Several pronounced current drops observed between the NTP2 and NTP0 monitors.}
        \label{fig:kick}
\end{figure}

There is a dedicated knob, which adjust settings of several correctors, to introduce the vertical orbit shift. This enables the introduction of extended vertical orbit distortion suitable for probing most of the technical region. Figure~\ref{fig:kick} presents combined TbT beam current measurements from all BPMs during the first sixteen turns following beam excitation. The vertical orbit was first shifted, so that the lifetime began to noticeably decrease. A pulsed kicker then excited coherent beam transverse oscillations around the new closed orbit. The data reveal an abrupt current drop between NTP2 and NTP0 monitors, indicating a potential aperture limitation in the section between them. For VEPP-4M, this section contains the RF cavity, insertable luminescent screen and shutters. Further localization requires targeted beam loss measurements with BLMs placed in the identified region.

When circulating beam is present, beam loss locations can be identified using local closed orbit distortions (bumps). At VEPP-4M, such orbit bumps can be created using sets of three or four dipole orbit correctors. During experiments, vertical orbit values were changed in several BPMs of the technical region, and corresponding corrector setting were applied. The beam lifetime was then recorded for different bump configurations and amplitudes.


\subsection{Electron beam loss measurements}

\begin{figure}[!th]
    \begin{center}
      \includegraphics[width=0.975\columnwidth,keepaspectratio]{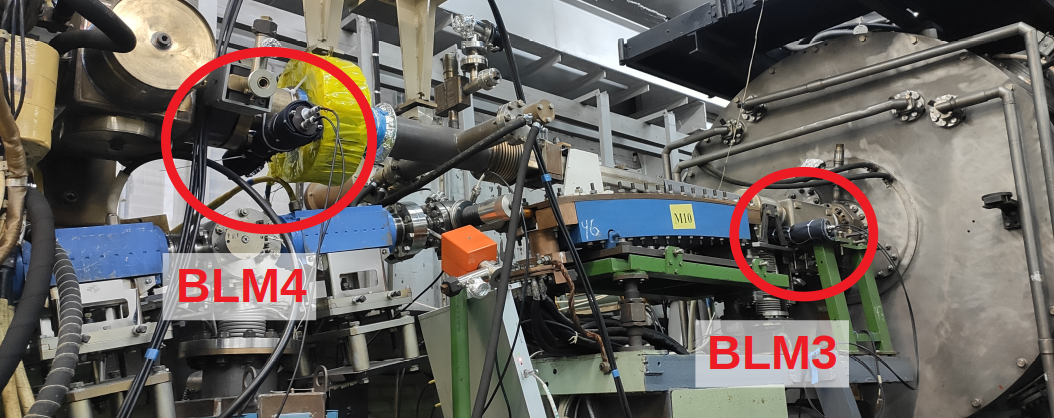}
    \end{center}
    \caption{Placement of BLMs in the VEPP-4M technical region.}
    \label{fig:photo}
\end{figure}

For further localization of the aperture limitation with higher precision, five portable scintillator-based BLMs were deployed in the VEPP-4M technical region. Figure~\ref{fig:layout} presents the schematic arrangement of these monitors along the accelerator lattice. The monitors were positioned under the vacuum pipe (Figure~\ref{fig:photo}). For better sensitivity, the center of the scintillator active volume is aligned with the center of the vacuum pipe. Numerical simulations show that the detector azimuthal position can be arbitrary. The bottom position was chosen, since it was planned to adjust the beam orbit primarily in the vertical plane due to the known lifetime asymmetry. The monitors were housed in a duralumin casing and wrapped in blackout fabric to protect against ambient light.
To monitor changes in the beam loss values and to compare several loss signals in real time, a digital form of displaying PMT signals was proved to be more practical. Therefore, it was proposed to record PMT signals using identical digital voltmeters. All BLM outputs were processed using the calibration procedure detailed in Subsection~\ref{subsection:calibr} and normalized by the beam current.


\begin{figure}[!th]
    \begin{center}
      \includegraphics[width=0.675\columnwidth,keepaspectratio]{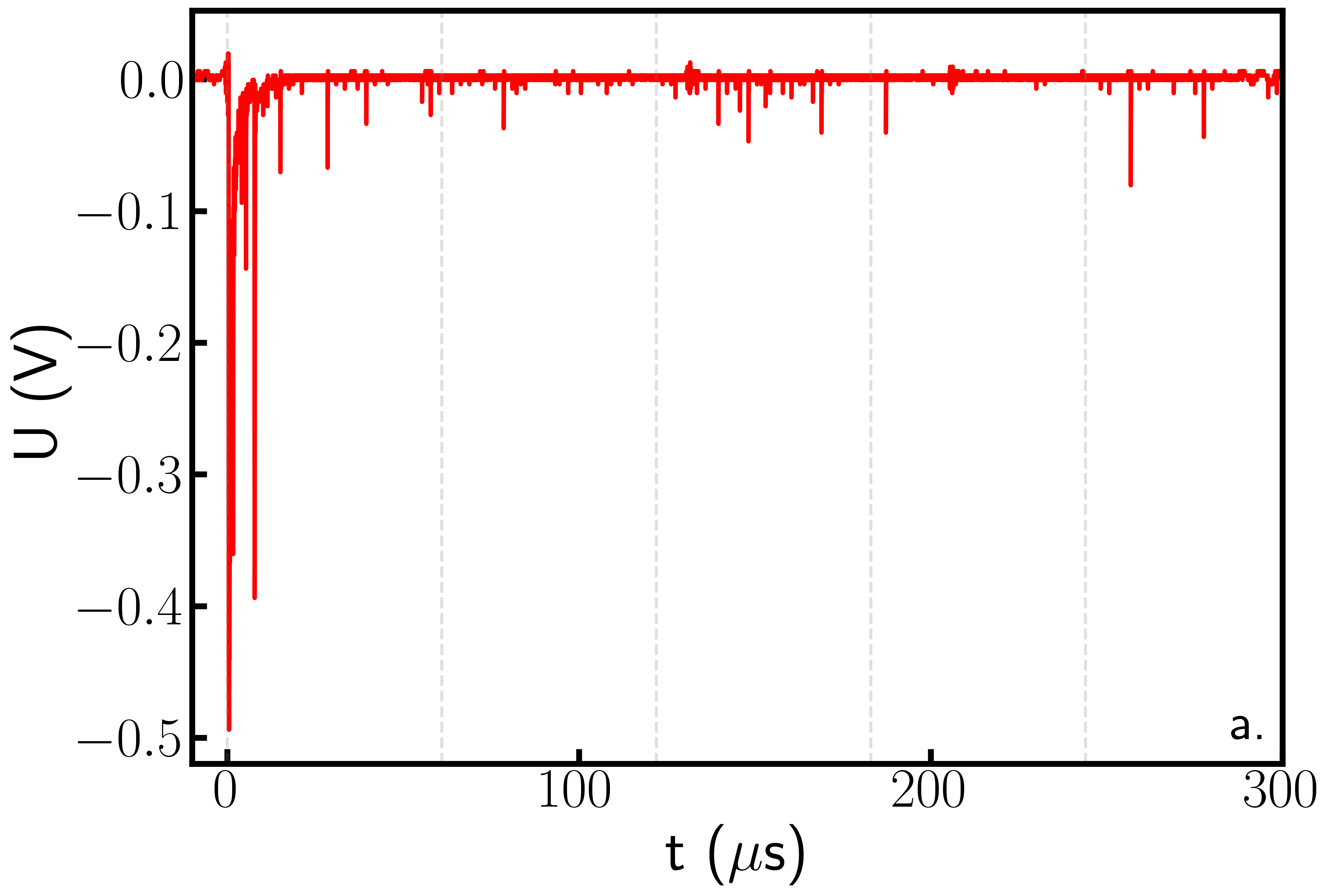}
      \includegraphics[width=0.30\columnwidth, height=0.325\textheight]{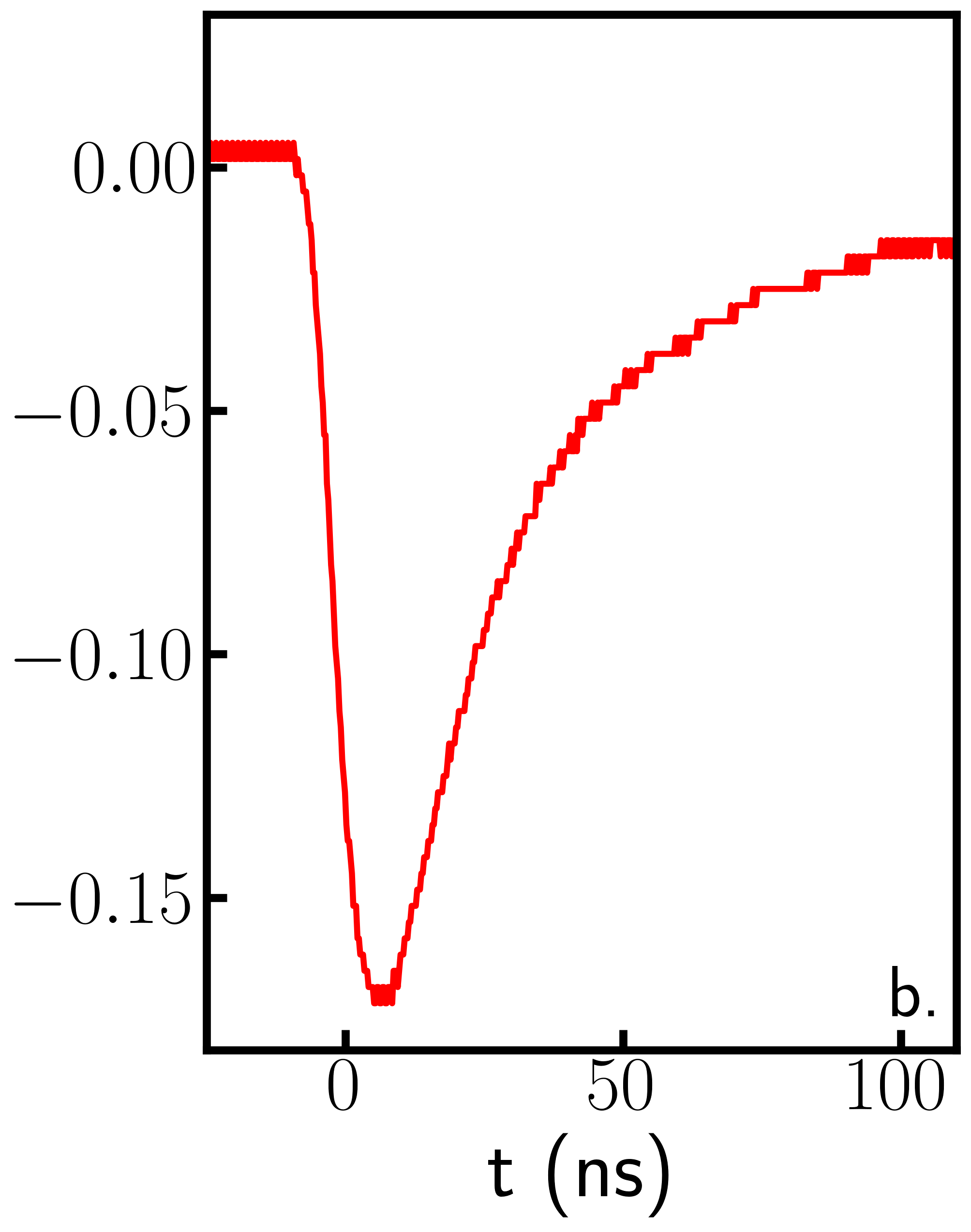}
    \end{center}
    \vspace{-0.75cm}
    \caption{a) TbT beam loss signals measured during injection at the VEPP-4M collider. The gray vertical lines indicate intervals corresponding to fifty revolution periods. b) Beam loss waveform recorded during one beam revolution in the VEPP-4M ring.}
    \label{fig:tbt_loss}
\end{figure}

To verify the sensitivity of scintillator-based BLMs, preliminary tests were performed, and loss signal waveforms were acquired. Measurements were carried out using one of the BLMs installed in the VEPP-4M technical region. Measurements were performed in both regular beam storage and injection modes. Figure~\ref{fig:tbt_loss}a shows PMT signal (negative polarity) for TbT beam losses during the first 250 revolutions of the injected beam, with a revolution period of \SI{1.22}{\micro\second}. Maximum losses are observed at the moment of injection. The signal amplitude gradually decreasing as beam oscillations damp. Figure~\ref{fig:tbt_loss}b shows the temporal structure of a single-turn beam loss for a \SI{1.9}{\giga\electronvolt} stored beam with a nominal orbit. Under these measurement conditions, the beam lifetime was approximately 2.5 hours with a beam charge of \SI{5}{\nano\coulomb}. The results confirm the monitor's sufficient sensitivity to detect and resolve individual TbT losses.


\begin{figure*}[!t]
    \begin{center}
      \includegraphics[width=\textwidth,keepaspectratio]{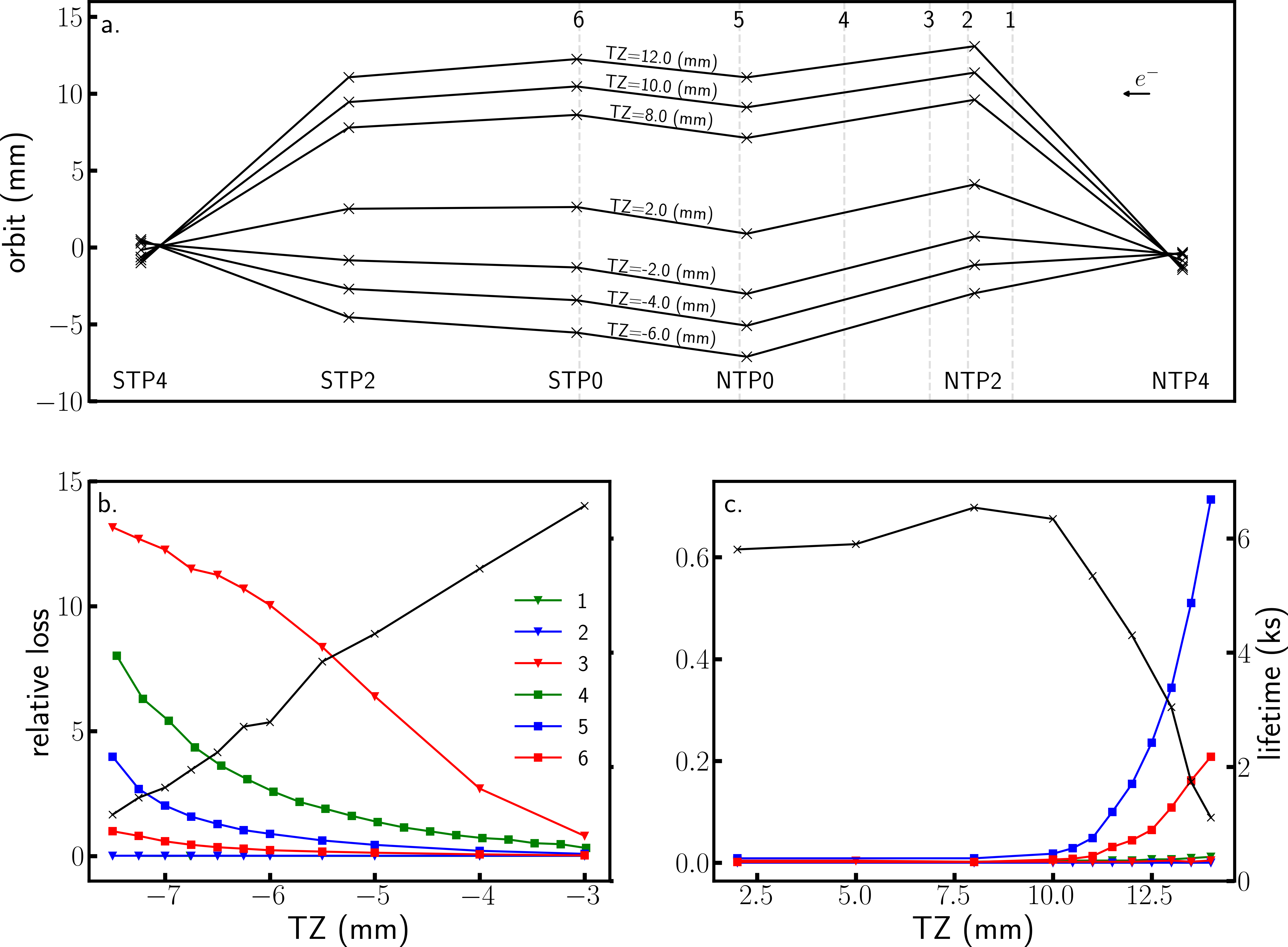}
    \end{center}
    \caption{a) Orbit distortion in the technical region. b) Beam losses (in color) and beam lifetime (in black) as functions of vertical orbit distortion for downward displacement. c) Corresponding measurements for upward displacement (BLM4 was not operational during these measurements).}
    \label{fig:tz}
\end{figure*}

Beam loss measurements were performed for different localized closed orbit distortions (Figure~\ref{fig:tz}a). Figures~\ref{fig:tz}b and \ref{fig:tz}c present the electron beam loss distribution as a function of the vertical orbit distortion in the technical region. The observed lifetime asymmetry with respect to the direction of the beam vertical orbit change reveals an aperture limitation at the bottom of the vacuum chamber. Losses begin to appear at $\textrm{TZ} = -3$ (where TZ denotes the vertical orbit displacement across the entire technical region) during downward orbit shift, accompanied by a notable reduction in beam lifetime. Peak losses are observed near BLM3 and progressively decrease in amplitude for downstream monitors BLM4 through BLM6.

Numerical simulations suggest that the observed monotonic decrease in loss signal magnitude corresponds to a single localized beam loss location. The scintillator-based BLMs exhibit exceptional detection sensitivity, resolving the extending exponential tail of the secondary particle distribution. This spatial extension, spanning multiple meters along the beamline, results from multiple backscattering of secondary particles within the vacuum pipe.

Figure~\ref{fig:tz}b shows no loss signals in BLM1 and BLM2 monitors, indicating the aperture restriction lies after these monitors with respect to the beam direction. This observation agrees with simulation results predicting negligible backward propagation of secondary particles. The simulated longitudinal loss distribution localizes the beam impact point between BLM2 and BLM3 monitors, which is consistent with the BPM measurements, indicating a potential aperture limitation in the section between NTP2 and NTP0 monitors. The linear growth of BLM3 loss signal compared to the exponential behavior for the monitors suggests the PMT has reached saturation, indicating the true loss rate exceeds the measured values at large orbit distortions.

Figure~\ref{fig:tz}c, for the upward orbit distortion, shows comparable losses emerging only around $\textrm{TZ} = +10$. The spatial loss distribution shows maximal loss signals near BLM5, while BLM1--BLM3 register no practically losses (BLM4 was not operational during these measurements). This asymmetric response relative to downward distortion case indicates a different beam loss position eventually reaching the vacuum pipe aperture, specifically, near the TL1.N and TL2.N magnets instead of RES3 resonator (Figure~\ref{fig:layout}). The observed TZ threshold difference ($\textrm{TZ} = +10$ versus $\textrm{TZ} = -3$) confirms a significant vertical asymmetry in the available aperture.


\begin{figure*}[!t]
    \begin{center}
      \includegraphics[width=\textwidth,keepaspectratio]{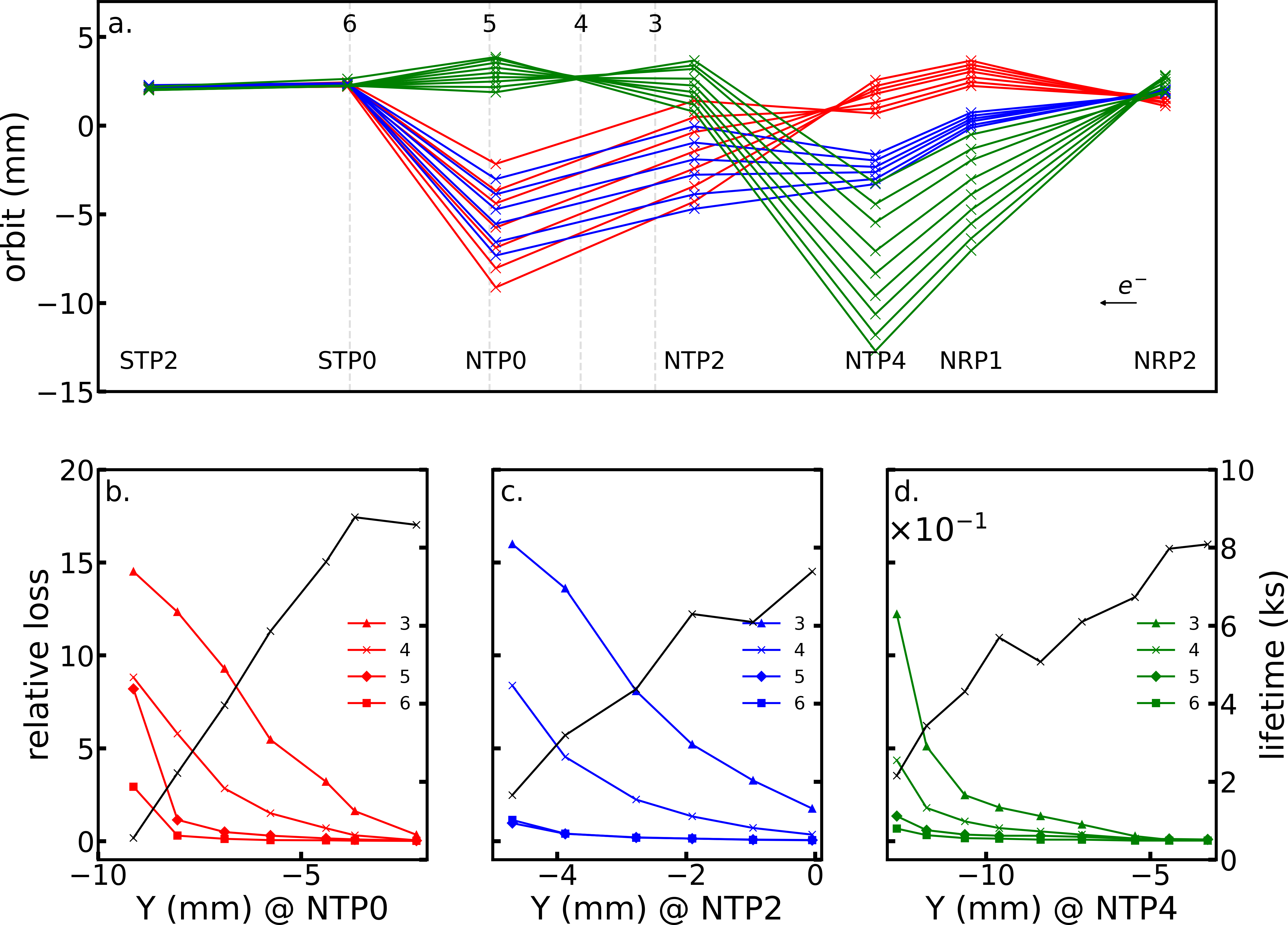}
    \end{center}
    \caption{a) Vertical orbit bumps in the technical region. Measured beam loss distribution (in color) and beam lifetime (in black) as functions of bump amplitude for: b) NTP0 bump (red), c) NTP2 bump (blue), and d) NTP4 bump (green).}
    \label{fig:bump}
\end{figure*}

Another set of measurements, performed using local vertical bumps, confirmed the hypothesis regarding the loss location. Figure~\ref{fig:bump} presents the electron beam loss distribution as a function of the vertical bump amplitude in the technical region, along with the induced closed orbit bumps. While both bumps at NTP0 and NTP2 have minimal orbit at the NTP0 location due to the lack of vertical correctors, the beam loss maximum occurs immediately downstream of NTP2. In the case of the bump at NTP4, the BLMs register a comparatively low loss level, suggesting that the primary beam loss occurs near NTP4. This loss location distance from the BLM positions results in only a minor fraction of secondary particles being detected.


\begin{figure}[!th]
    \begin{center}
      \includegraphics[width=0.975\columnwidth,keepaspectratio]{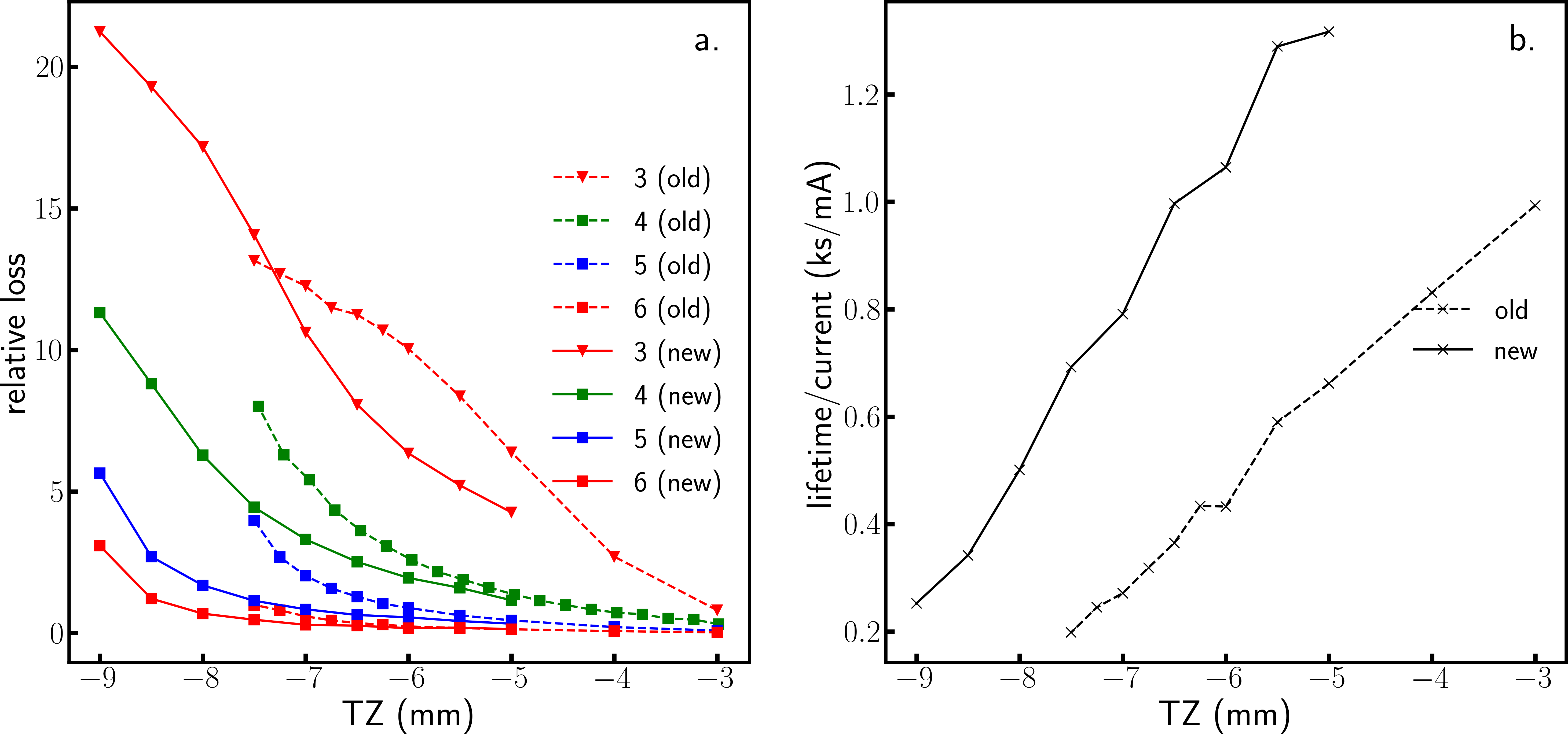}
    \end{center}
    \vspace{-0.75cm}
    \caption{a) Beam losses as functions of vertical orbit distortion before (dashed) and after (solid) the BPM geodetic displacement settings update and b) corresponding normalized beam lifetime.}
    \label{fig:geo}
\end{figure}

Having identified the loss location between monitors BLM2 and BLM3, we conducted a detailed inspection of this region for potential aperture restrictions and possible obstacles origin. The RES3 resonator located between these monitors contains multiple vacuum shutters that could potentially introduce aperture limitations. However, radiographic analysis of the resonator excluded this possibility. Subsequent investigation revealed inaccuracies in the geodetic displacements of certain BPMs, particularly NTP2, within the technical region. After applying updated BPM geodetic displacement settings, the observed asymmetry in the beam lifetime due to the direction of closed orbit vertical distortion was significantly reduced (Figure~\ref{fig:geo}). These results demonstrate the practical ability of the proposed method for identifying aperture limitations, in particular, limitations caused by inaccurate geodetic information.


\begin{figure}[!th]
    \begin{center}
      \includegraphics[width=0.7\columnwidth,keepaspectratio]{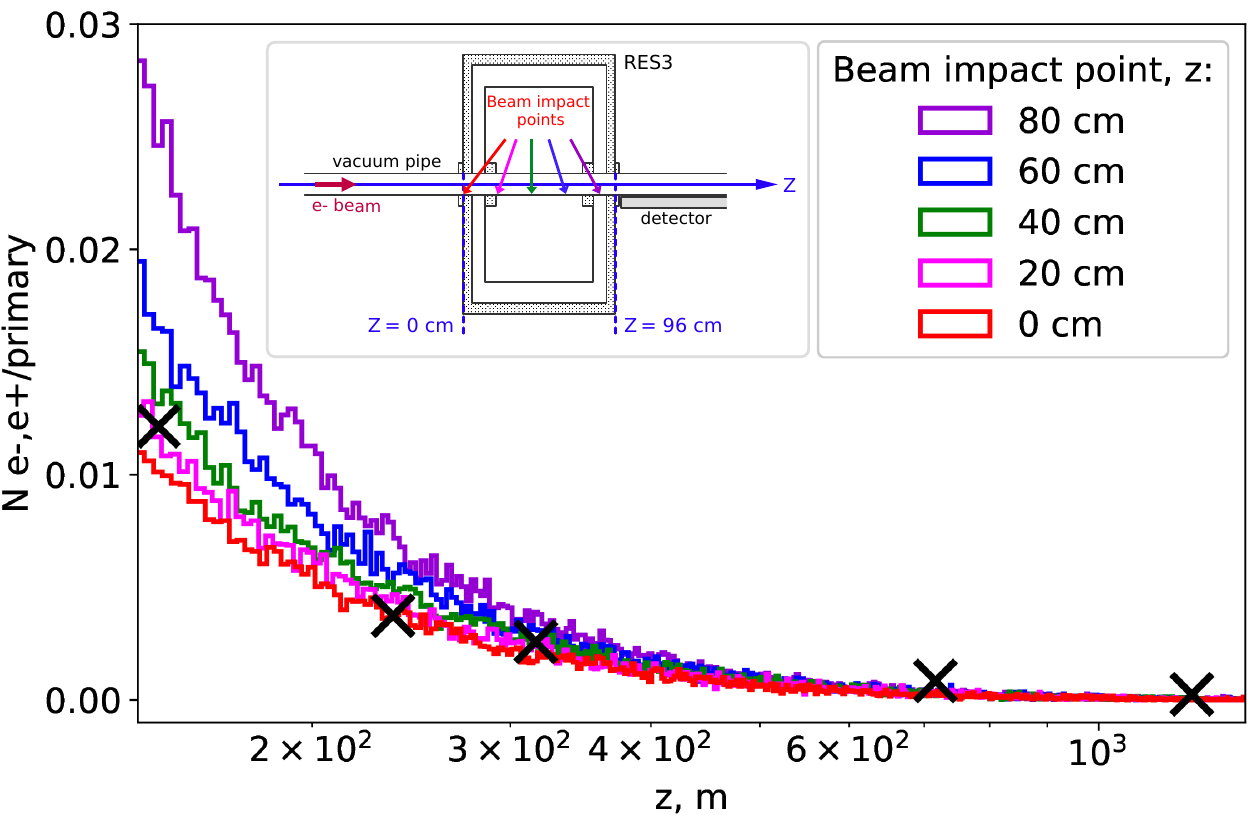}
    \end{center}
    \caption{Comparison of beam simulated loss distribution tails for different beam loss locations inside the RES3 resonator (in color) with matched measured results (black crosses).}
    \label{fig:comparison}
\end{figure}

After experimentally localizing the beam loss region, we performed FLUKA simulations of beam losses within the RES3 resonator and compared the results with the discrete secondary particle distribution sampled by the BLM system. Figure~\ref{fig:comparison} presents a comparison between FLUKA simulations for different potential beam impact points inside the resonator and actual BLM measurements acquired at $\textrm{TZ} = -5$ (without PMT saturation). To enable quantitative comparison between experimental data and simulations, normalization of the measured data was required. The normalization factor was optimized to achieve the best match with the characteristic exponential tail of the simulated loss distribution. The experimental results show optimal agreement with simulations corresponding to beam losses occurring in the first half of the \SI{1}{\meter}-long RES3 resonator (at $z = \SI{0}-\SI{40}{\centi\meter}$). The simulation demonstrates that FLUKA modeling can provide enhanced spatial resolution for precise beam loss localization beyond the resolution of the BLM measurements.


\begin{figure}[!th]
    \begin{center}
      \includegraphics[width=0.7\columnwidth,keepaspectratio]{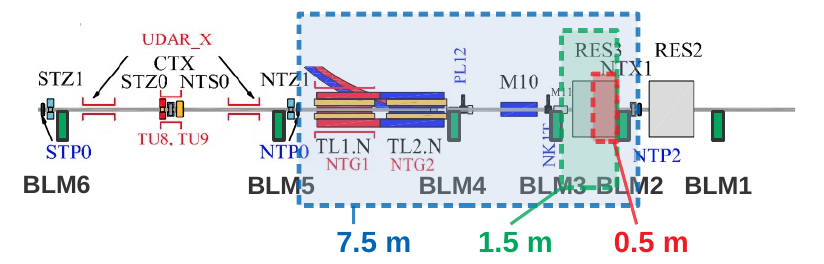}
    \end{center}
    \caption{Aperture limitation region localized by BPMs (blue), BLMs (green), and FLUKA simulations (red).}
    \label{fig:regions}
\end{figure}

The combined method utilizing TbT BPM and BLM measurements successfully identified and localized an aperture restriction in the VEPP-4M technical region. Figure~\ref{fig:regions} illustrates the sequential refinement process of the beam loss localization. The initial localization, performed using TbT BPM data, localized the loss region to a \SI{7.5}{\meter} section between two BPMs. The BLM measurements further refined the loss position to a \SI{1.5}{\meter} region around the RES3 resonator. The final localization, achieved through extrapolation based on the matching between simulated beam loss distribution tails and BLM measurements, pinpointed the loss location to \SI{0.5}{\meter} entrance segment of the RES3 resonator.


\section{Summary}
\label{section:summary}


In this work, we presented a method for identifying aperture limitations in circular accelerators by combining TbT BPM and BLM measurements. The proposed technique is applicable during critical operational stages in charged particle accelerators, such as initial machine commissioning or normal operation when beam circulation is established. The approach employs TbT beam current measurements from BPMs to identify initial regions of aperture restriction by observing sudden current drops between adjacent BPMs. Subsequent refinement is achieved using beam loss data from portable scintillator-based BLMs. By repositioning BLMs along the accelerator lattice, sub-meter spatial resolution in loss localization can be attained. To enable loss signal comparison across different detectors, the BLM system should first undergo relative sensitivity calibration before measurements are taken.

Experimental validation was conducted at the VEPP-4M collider, where an aperture restriction was suspected in the technical region localized by TbT BPM current data analysis. For precise localization, five identical scintillator-based BLMs were deployed. Relative sensitivity calibration of the detectors was performed using a strontium-90 radioactive source, enabling quantitative comparison of loss signals. To correlate measured BLM responses with physical aperture constraints, the FLUKA simulations were performed. These simulations modeled beam loss distributions under different aperture limitation scenarios. The simulated and experimental beam loss results exhibited strong agreement, demonstrating that FLUKA modeling can enhance spatial resolution beyond the intrinsic limitations of standalone BLM measurements.

The combined method successfully identified an aperture restriction in the VEPP-4M's technical region, attributed to inaccurate geodetic alignment of BPM centers. The integration of preliminary BPM-based localization with BLM measurements and Monte Carlo simulations establishes a robust framework for resolving aperture limitations with sub-meter precision.


\section*{Declaration of competing interest}

The authors declare that they have no known competing financial interests or personal relationships that could have appeared to
influence the work reported in this paper.

\section*{Acknowledgments}

This work was partially supported by the Ministry of Science and
Higher Education of the Russian Federation (project FWUR-2025-0004).




\end{document}